# HRMOS: a very high-resolution, multi-object spectrograph for the ESO VLT

Andrea Bianco[a], Sofia Randich[b], Emma Fernandez Alvar[c], Oscar Gonzalez[d], Sérgio Sousa[e], Thomas Bensby[f], Letizia Caito[g], Enrico Giro[h], Laura Magrini[b], Marco Riva[a]

[a]INAF–Osservatorio Astronomico di Brera, Via E. Bianchi 46, 23807 Merate (LC), Italy, [b]INAF–Osservatorio Astrofisico di Arcetri, Largo Enrico Fermi 5, 50125 Firenze, Italy; [c]Universidad de La Laguna, Calle Padre Herrera, s/n, 38200 La Laguna, Santa Cruz de Tenerife, Spain; [d]UK Astronomy Technology Centre, Royal Observatory Edinburgh, Blackford Hill, Edinburgh EH9 3HJ, UK, [e]Instituto de Astrofísica e Ciências do Espaço, Rua das Estrelas, 4150-762 Porto, Portugal; [f]Lund Observatory, Lund University, Box 43, SE-221 00 Lund, Sweden, [g]INAF, Viale del Parco Mellini 84, 00136 Roma, Italy; [h]INAF-Osservatorio Astronomico di Trieste Via G.B. Tiepolo, 11, 34143 Trieste, Italy

## ABSTRACT

HRMOS (High-Resolution Multi-Object Spectrograph) is a proposed new instrument for the ESO Very Large Telescope (VLT) developed in the context of the ESO VLT Beyond 2030 call. It is designed to fill a unique and currently unoccupied region in the observational landscape: the combination of very high spectral resolution (R = 80,000) with multi-object capability (50–60 simultaneous targets), a radial-velocity (RV) precision of 10 m s$^{-1}$, and coverage of three key spectral windows (385–421 nm, 480–522 nm, 623–677 nm). Scientifically, HRMOS will address a rich portfolio of high-priority astrophysical questions spanning from f giant exoplanet formation, to nucleocosmochronology and constraints on cosmological parameter, to probing hierarchical galaxy assembly outside the Milky Way . The instrument is based on a modular approach and it consists in four primary subsystems, the Front End with a hybrid fiber-positioning and atmospheric dispersion correction (ADC) architecture, the Fiber Link with double-scrambling and the image slicers, three spectrographs based on volume phase holographic (VPH) gratings covering the three spectral ranges and a Calibration Unit.

## 1. INTRODUCTION

The past three decades have witnessed a revolution in ground-based spectroscopic surveys driven by multi-object spectrographs such as RAVE, Gaia-ESO[1], APOGEE[2], GALAH, and LAMOST[3]. A new generation of facilities, i.e. 4MOST[4], WEAVE[5], MOONS[6], and the proposed Wide-field Spectroscopic Telescope (WST)[7] will extend this reach to much larger samples at resolving powers up to R≈ 20,000 – 40,000.

Despite this anticipated progress, the international landscape lacks an instrument on an 8-metre-class telescope that combines the statistical power of MOS with very high spectral resolution (see Fig. 1). Single-object spectrographs such as ESPRESSO[8], UVES, HARPS[9], and the forthcoming ELT/ANDES[10] provide exquisite resolution and RV precision for individual stars, but no facility can deliver this simultaneously for large statistical samples. HRMOS is conceived precisely to close this gap.

*andrea.bianco@inaf.it; www.hrmos.eu

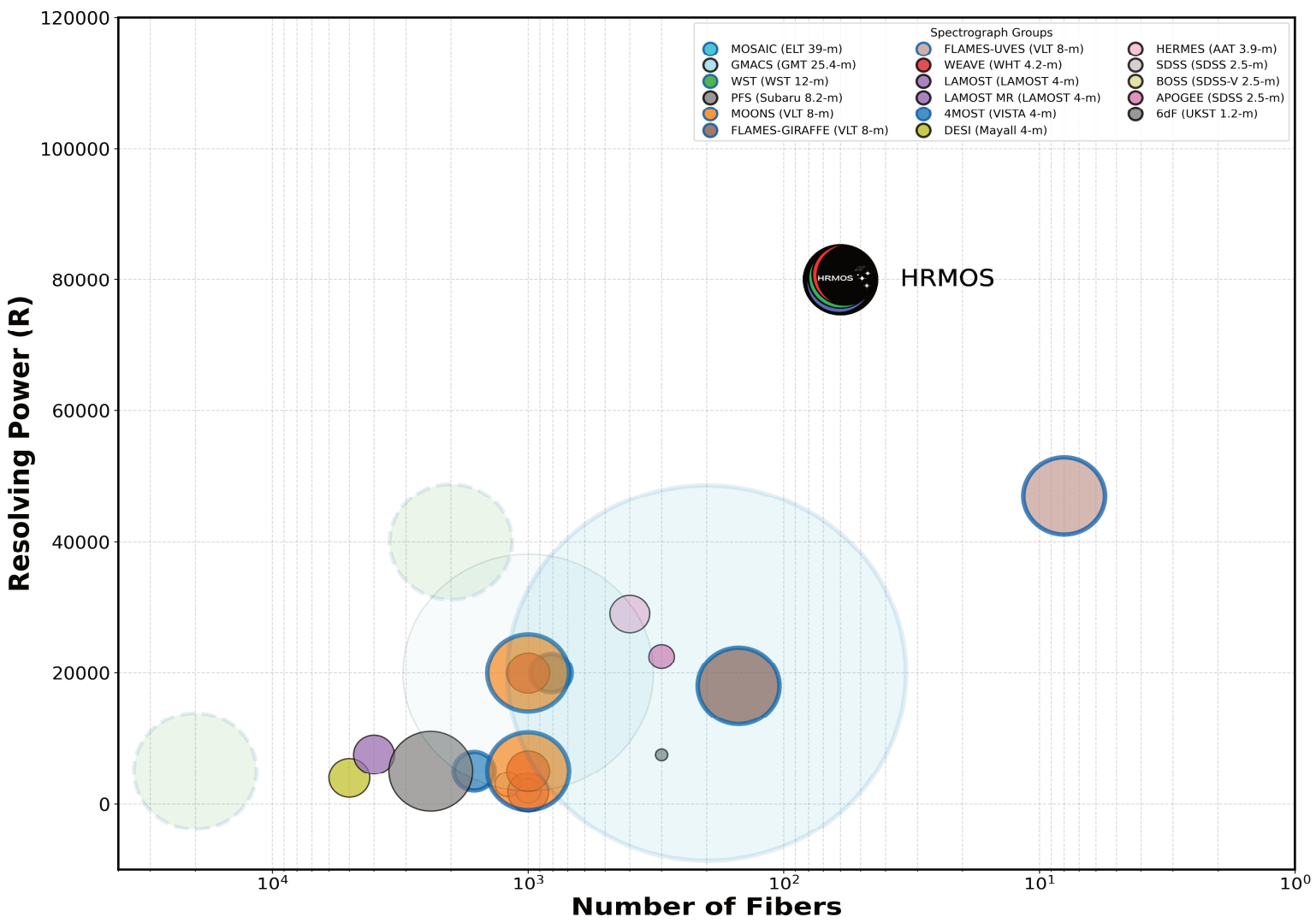

Figure 1. HRMOS in the context of multi-object spectrographs. The resolving power R is shown as a function of the number of fibers, with the symbol sizes proportional to the collecting area of the corresponding telescope. HRMOS is represented by its logo, positioned at $R \approx 80{,}000$ and a multiplex of 60 fibers (within the range 50–60). Future instruments are represented with transparent circles; proposed facilities are denoted with dashed contours, while ESO instruments are highlighted with blue contours.

The HRMOS initiative translates into a consortium of almost 200 researchers and engineers. The instrument concept has evolved from an initial design to a solid system-level concept, and the science case has been thoroughly documented in the HRMOS White Paper v2[11]. This proceedings paper summarizes the primary science drivers and the current instrument design that will be submitted in the context of the ESO VLT Beyond 2030 call[12].

## 2. PRIMARY SCIENCE CASES

HRMOS addresses a broad portfolio of astrophysical science cases that require, simultaneously, high spectral resolution and large statistical samples. Some important scientific drivers are outlined below[11].

### 2.1 Exoplanets across Galactic Environments

Observationally probing how exoplanetary system properties depend on the chemical and dynamical environment is one of the most pressing open questions in planetary science, that would allow new constraints on models of planet formation Whilst giant planet occurrence rates correlate with host-star metallicity, the role of the broader environment, stellar density, radiation fields, remains poorly constrained due to the limited diversity of environments explored so far.

HRMOS will enable systematic RV surveys of stars in open clusters, globular clusters, the Galactic bulge, and nearby dwarf galaxies. With radial velocity (RV) precision of 10 m $s^{-1}$ and 50–60 simultaneous targets, HRMOS will be at least 500 times more survey-efficient than ESPRESSO for the same number of targets. Key questions include: how do environment, metallicity, and stellar multiplicity affect the formation and occurrence of giant planets? How does stellar evolution shape the demographics of short-period planets?

### 2.2 Origin of the Elements and Nucleosynthesis

The origin of some elements in the periodic table, such as carbon, sulphur and -crucially- elements heavier than iron, produced through rapid and slow neutron-capture processes (r- and s-process), remains a fundamental open problem. Measuring tiny lines and isotopic abundance ratios—such as $^{12}C/^{13}C$ and barium isotopes, requires both very high spectral resolution and high signal-to-noise ratio (SNR). HRMOS will deliver for the first time these measurements for large samples of stars in clusters spanning the full range of Galactic metallicities and ages, constraining the sites, rates, and yields of these elements, in particular those from neutron stars  mergers and kilonovae.

### 2.3 Hierarchical Formation and Evolution of Nearby Galaxies

The assembly history of the Milky Way and its satellites is encoded in the detailed and precise chemical abundances and kinematics of individual stars. HRMOS will enable large-scale, high precision surveys of stellar populations in nearby dwarf galaxies (Sagittarius, Magellanic Clouds ). Its crowded-field capability (minimum fiber separation of 10″ in the central zone) is particularly valuable for targets in the dense central regions of these systems, inaccessible to any current facility.

### 2.4 Nucleocosmochronology

The ages of the oldest Milky Way stars provide a direct lower bound on the age of the Universe, allowing an independent estimate of cosmological parameters and, in particular, $H_0$The nucleocosmochronology technique uses the radioactive decay of $^{232}$Th relative to stable r-process elements such as Eu to derive model-independent stellar ages. The precise measurement of the extremely of the weak Th absorption lines requires R$\gtrsim$ 80,000 and very high SNR. HRMOS will enable, for the first time, nucleocosmochronological age determinations for statistically significant samples of several globular cluster members.

### 2.5 High-Resolution Spectroscopy of Star Clusters

Star clusters are fundamental laboratories for stellar physics, chemical evolution, and dynamics. HRMOS will deliver simultaneous high-resolution spectra for large samples of cluster members, enabling studies of: (i) multiple stellar populations in globular clusters through abundance patterns of light elements and isotopic ratios; (ii) stellar activity and Doppler tomography in young clusters; (iii) precise internal kinematics and binary fractions; and (iv) diffusion, mixing, and lithium evolution across the HR diagram.

## 3. INSTRUMENT OVERVIEW AND TOP-LEVEL REQUIREMENTS

HRMOS is designed to be mounted at the Nasmyth focus of one of the VLT Unit Telescopes, exploiting the full 25-arcmin field of view. The instrument is designed as a significant upgrade of FLAMES, leveraging heritage from MOONS, KMOS, ESPRESSO, and ANDES (ELT). The top-level requirements (TLRs) are summarized in Table 1.

Table 1. HRMOS top-level requirements.

| Parameter | Requirement | Goal |
|---|---|---|
| Resolving power | R = 80,000 | – |
| Spectral range | 385–421 / 480–522 / 623–677 nm | – |
| RV precision | 10 m s$^{-1}$ | 5 m s$^{-1}$ |
| No. of fibres | 50 | 60 |
| Min. fibre separation | 10″ (inner) / 30″ (outer) | – |

The total spectral coverage is about 130 nm, crucially extending down to 385 nm. To be noted also the requirement of the vicinity of the objects in the central part of the FoV, which is 10” to probe crowded fields. In the outer part of the field, the separation is 30”, under the assumption that the outer regions of the clusters are less crowded. The block diagram of HRMOS is reported in Fig. 2.

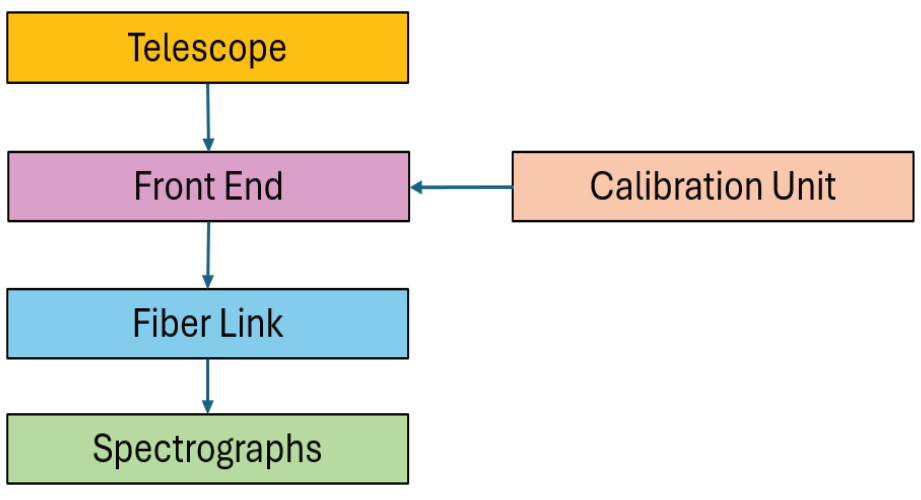

Figure 2. High-level block diagram of HRMOS. The instrument comprises four primary subsystems: Front End, Fiber Link, Spectrographs, and Calibration Unit.

The signal-to-noise performance is illustrated in Fig. 3. HRMOS will achieve SNR > 50 per resolution element in 1 hour for solar-type stars to G = 16.5 in the red arm and G = 15 in the blue arm.

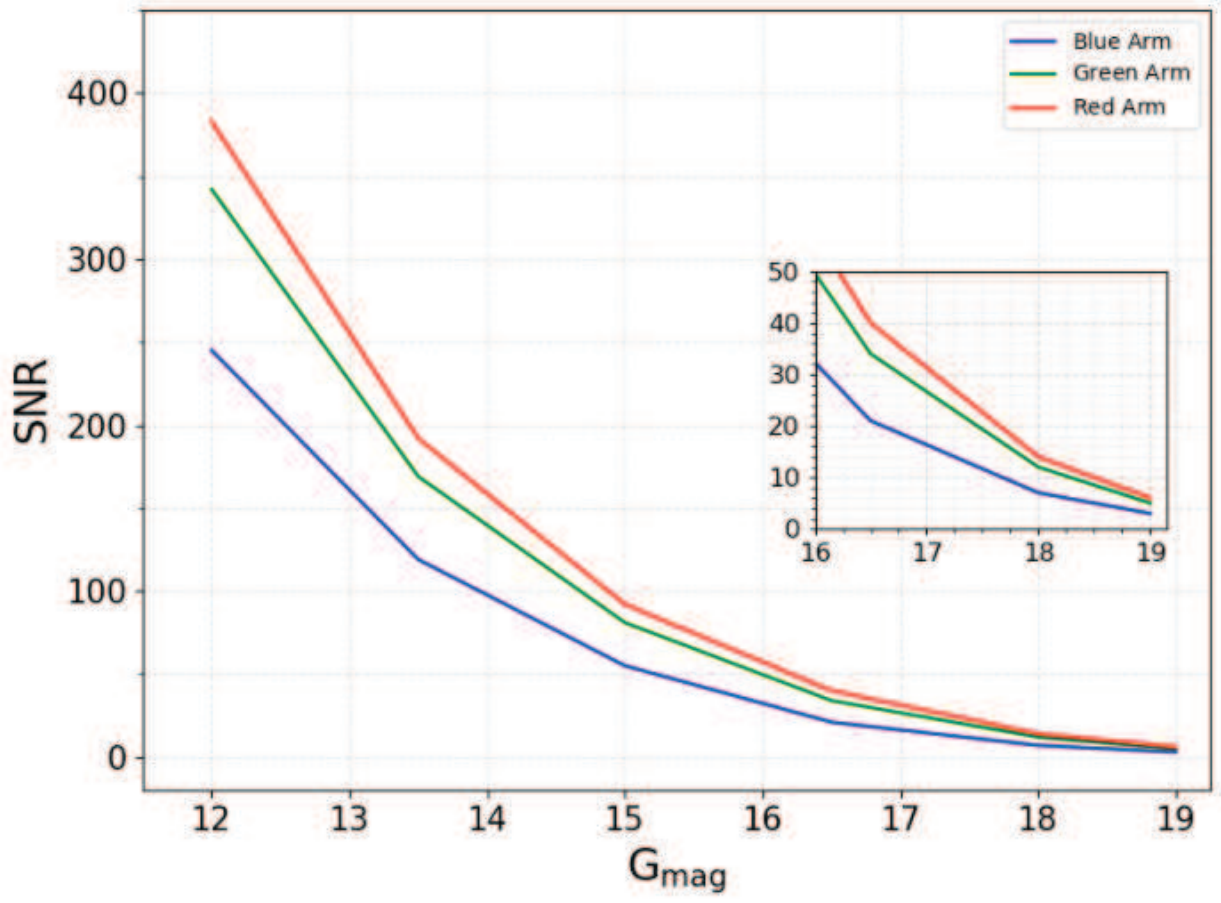

Figure 3. SNR per resolution element as a function of G magnitude for a 1-hour exposure (airmass 1, dark sky) across the three HRMOS spectral arms (blue, green, red)[11].

## 4. INSTRUMENT DESIGN

The HRMOS instrument consists of four primary subsystems. The overall design philosophy prioritizes RV stability, throughput, and crowded-field capability, drawing on proven heritage from existing high-precision and multi-object spectrographs. In this section, we describe the four items.

### 4.1 Front End

The Front End interfaces the VLT focal plane with the fiber link. Its core functions are: (i) collecting stellar light; (ii) correcting the atmospheric dispersion (ADC); (iii) fiber injection; and (iv) calibration light injection. It exploits the 25′ field of view via a MOONS-based field corrector on a rotating platform.

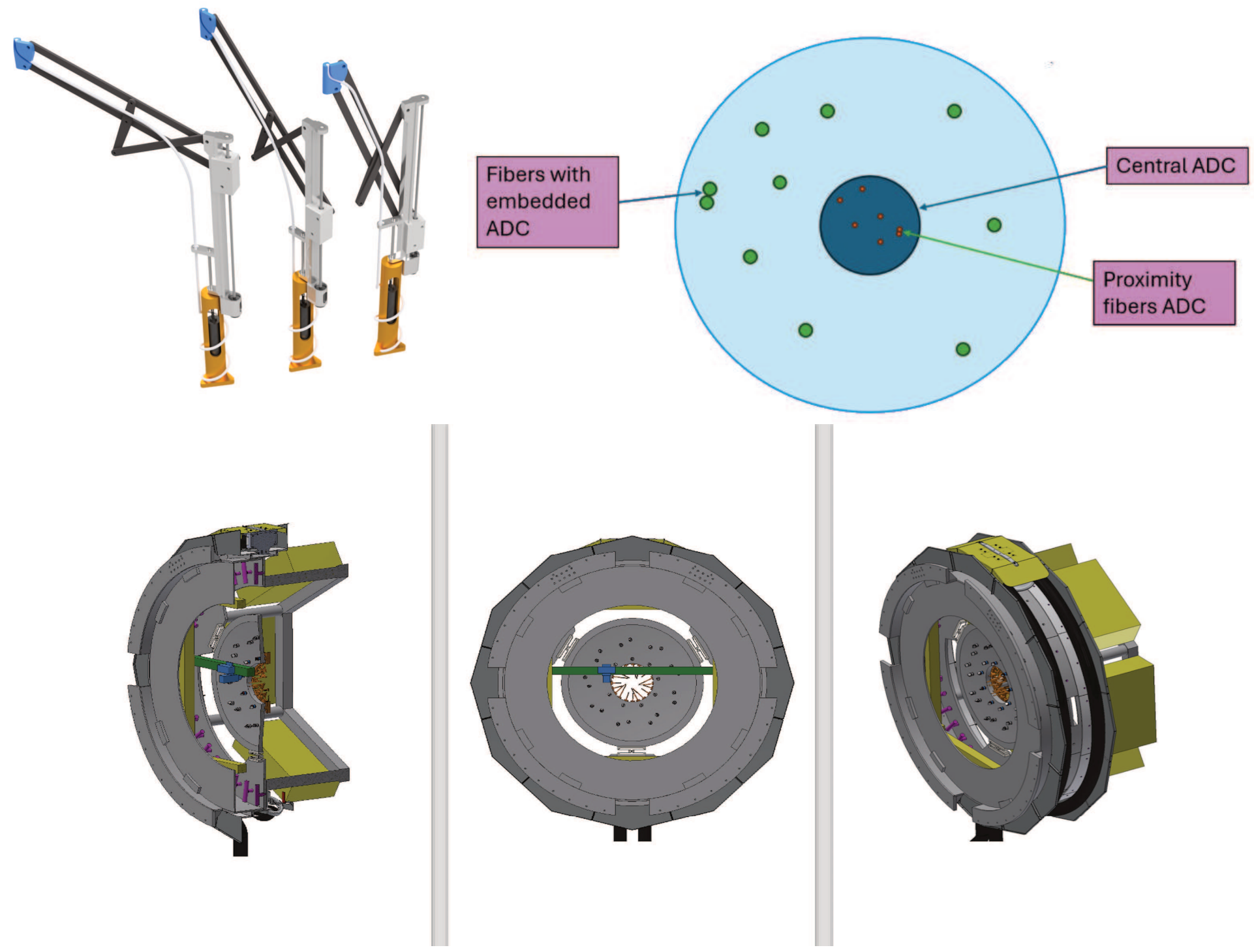

Figure 4. Top left: Scheme of the classical r–θ positioner; top right: diagram of the hybrid ADC system; bottom: CAD rendering of the HRMOS Front End mounted at the Nasmyth focus. The rotating structure, inherited from MOONS, hosts the fiber-positioning modules and the hybrid ADC solution.

Three fiber-positioning concepts were evaluated: a classical r–θ positioner (Fig. 4) coming from MOONS/KMOS heritage[6,13] and a novel pick-and-place system with integrated miniaturized ADCs. The first approach requires a large ADC in front of the fibers, the second shift the correction to micro ADC on the side of the front end. A third option that is evaluated is a hybrid two-zone solution (Fig. 4): a single shared ADC covers the central field (fibers ≥10″ apart), essential for globular cluster cores, while individual compact ADCs are integrated in the peripheral pick-off modules (≥30″). This could represent the best compromise between crowded-field capability, ADC performance, and technological readiness. A trade-off activity will be necessary to evaluate TRLs and risks and to achieve the most effective solution.

### 4.2 Fiber Link

The Fiber Link connects the Front End to the spectrograph arms via 50 (goal: 60) identical per-target modules, each performing: (i) light transmission; (ii) spectral splitting via dichroic beam-splitters; (iii) near- and far-field double-scrambling (suppression factor $\sim10^4$, reducing guiding-induced RV systematics to sub-cm $s^{-1}$); (iv) image slicing into 19 sub-pupils to match the telescope étendue to the spectrograph slit (2.7 pixels per resolution element); and (v) insertion of a simultaneous calibration signal (Sym Calib). The diameter of the fiber on sky corresponds to 1” (140 µm core diameter). Since a high resolving power of 80,000 is required, the fiber is sliced in coupled to 19 smaller fibers. This means a larger number of fibers and a longer pseudo slit on the spectrograph. We plan to develop small boxes, one for each sky fiber to separate the three spectral ranges and slice the 1” fiber in the 19 small fibers. At the end, this will result in 50 – 60 elements with 19x3 fibers each emerging from them. This idea comes from the proposal in MSE[14]. A scheme of this approach is reported in Fig.5.

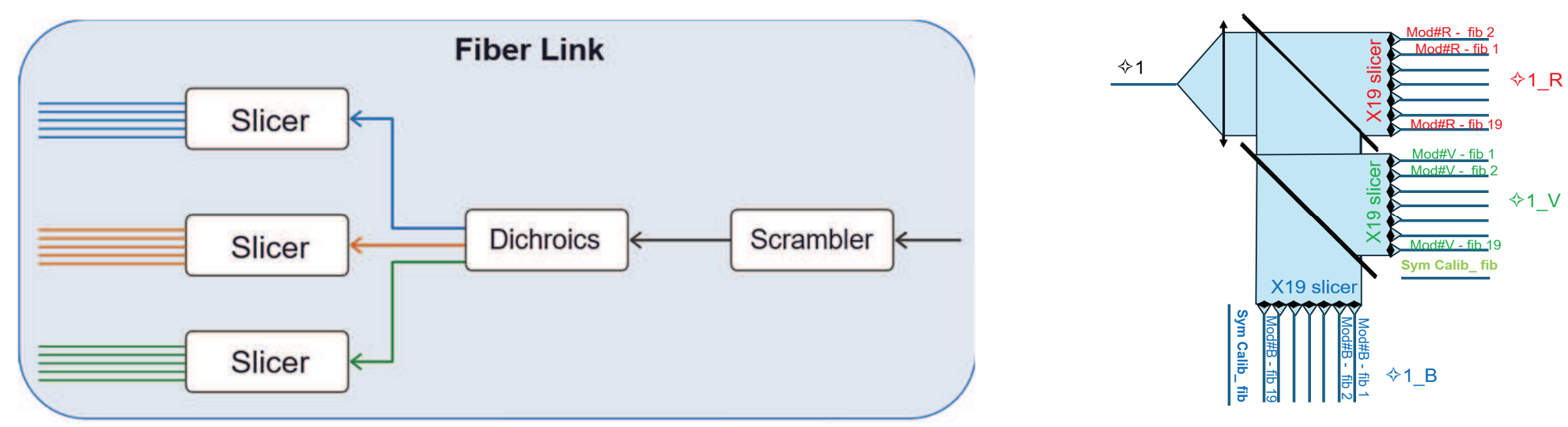

Figure 5. Conceptual scheme of the HRMOS fiber link. Each of the 50 (goal: 60) per-target modules contains a double scrambler, three dichroic beam-splitters (B/V/R channels), three ×19-slice slicer assemblies, and a simultaneous calibration fiber (Sym Calib) per channel.

Concerning the design of fiber link components and technologies (Fig. 6), they draw on the expertise that is building up in the framework of ELT ANDES project will be exploited in HRMOS.

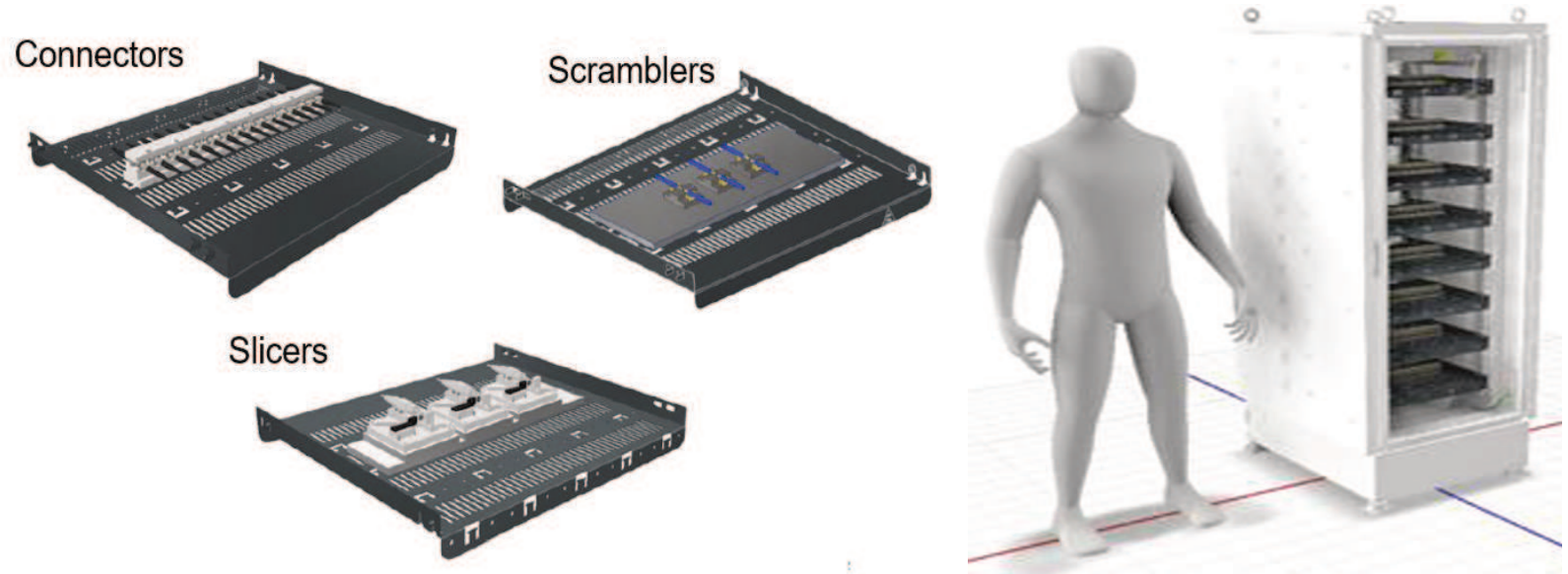

Figure 6. Fiber-link hardware: Front-End assembly trays with connectors, scramblers and slicers (left), and full fiber-link rack housing all 50 per-target modules (right). Human figure for scale (credits ANDES).

### 4.3 Spectrographs

The spectrograph component benefits of the fiber link approach and, since the spectral bands are already separated, three spectrograph arms, blue (B: 385–421 nm), green (V: 480–522 nm), red (R: 623–677 nm) are expected sharing a common optical design housed in a single temperature- and pressure-stabilized enclosure. The pressure and temperature control is necessary to achieve the target precision in the radial velocity. In particular, thermal stability of a few tenth of mK $h^{-1}$ is required, consistent with HARPS and ESPRESSO heritage.

Each arm performs: pseudo-slit formation consisting in 19 slices, plus a calibration fiber and an “empty fiber” (to separate each object) stacked per target, collimation; dispersion by a VPH grating[15], and detection on a 2 × (9k × 9k, 10 µm pixel) CCD/CMOS mosaic at F/2.96. The spectral sampling is 2.7 pixels per resolution element.

The actual optical design has a collimated beam with a diameter of 420 mm and VPHGs working at 41° of AOI (Fig. 7. The expected diffraction efficiencies are reported in Fig. 7. They increase going from blue to red with an average efficiency of about 70% - 75%.

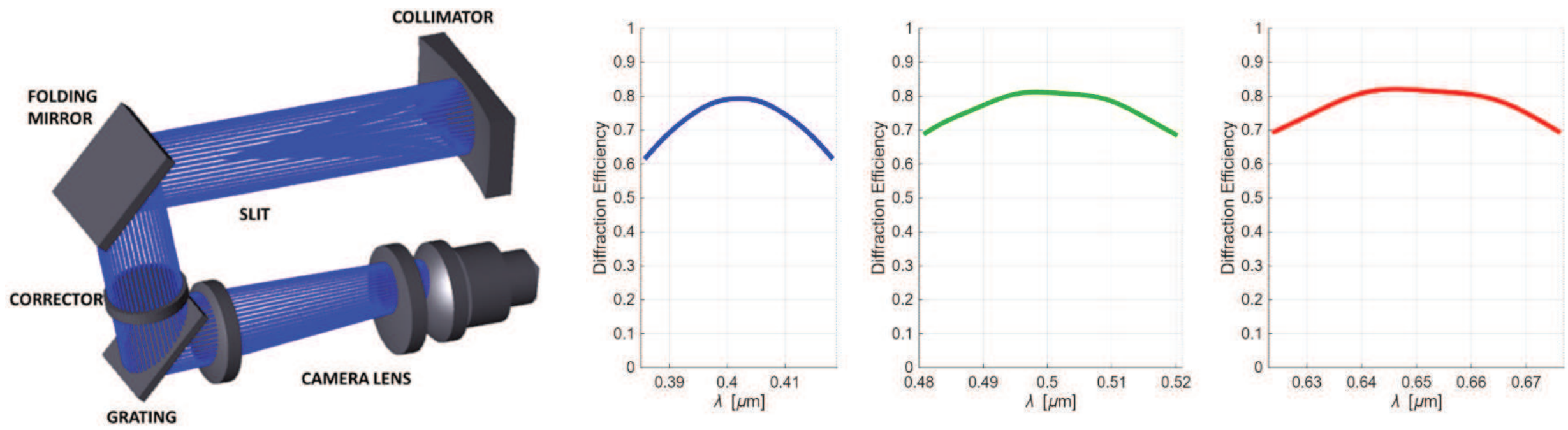

Figure 7. Left-hand panel: spectrograph design; right-hand panel: the expected diffraction efficiency for the three gratings considering the production losses.

Another important aspect is the selection of the detector, since CCD technology is making way to CMOS technology and strong efforts are going in this direction. We plan to be part of this activity enabling the use of CMOS detectors in HRMOS. Globally the spectrographs will have a volume ~$3.2 \times 2.0 \times 1.0$ m$^3$ as shown in Fig. 8.

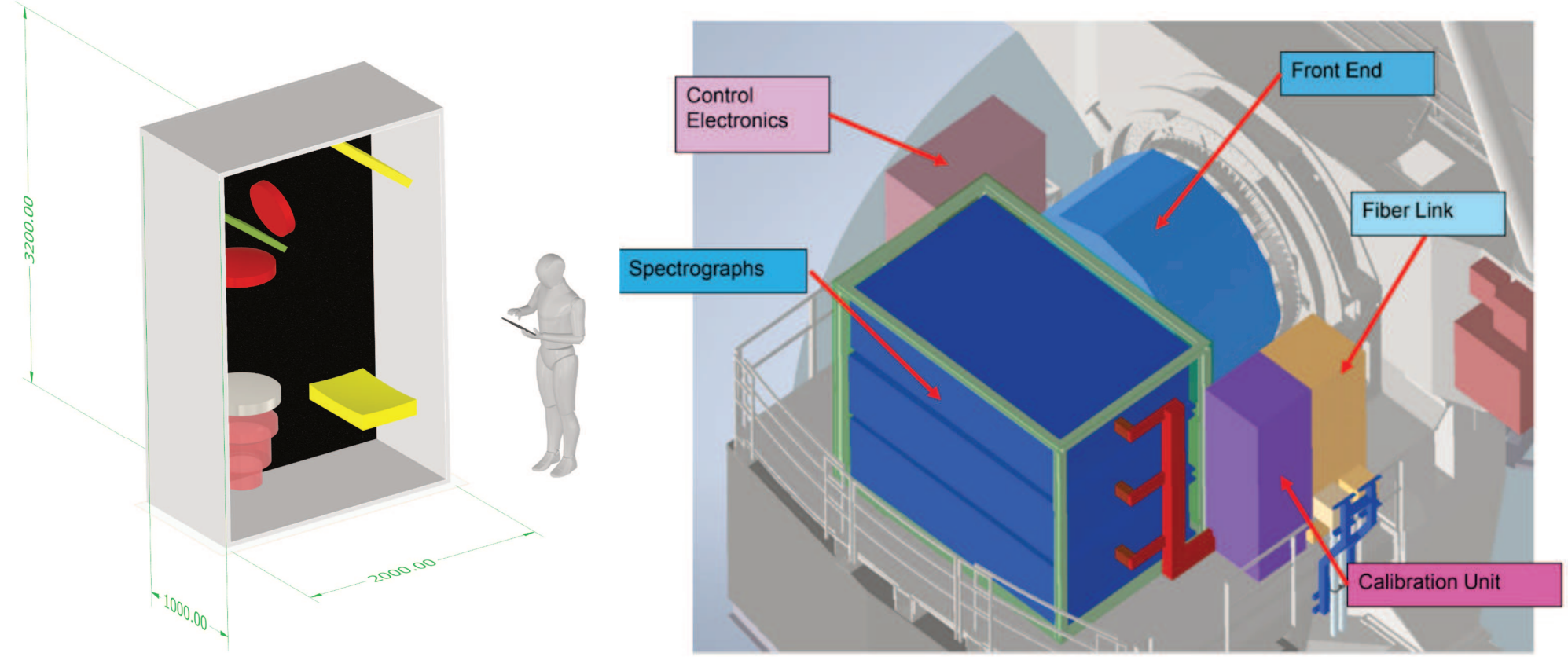

Figure 8. Three-dimensional CAD model of one HRMOS spectrograph arm ($3.2 \times 2.0 \times 1.0$ m$^3$). Human figure for scale. On the right the layout of the Nasmith platform hosting the HRMOS system

### 4.4 Calibration Unit

The Calibration Unit provides wavelength calibration, flat-field illumination, and simultaneous RV reference signals. A Laser Frequency Comb (LFC) is the baseline simultaneous reference for the RV programme, injected into dedicated Sym Calib fibres that run alongside each science fiber through the scrambler and slicer stages to the detector. Fabry–Pérot etalons and hollow-cathode lamps serve as alternative or complementary sources for daytime calibrations.

### 4.5 Radial-Velocity Error Budget

An important point is the evaluation of the error budget in the RV for the exoplanet science case. The dominant systematic contributions to the RV error budget are thermo-mechanical drifts of the spectrograph bench and detector effects. When calibrated with the simultaneous LFC reference (assuming 30% residuals, a conservative estimate), the total estimated RV error is 8.2 m s$^{-1}$, comprising a random component of 1.4 m s$^{-1}$ and a systematic component of 6.7 m s$^{-1}$, comfortably within the 10 m s$^{-1}$ requirement. Such budget has been calculated using same approach employed in dedicated spectrographs such as ESPRESSO and HARPS [8,9]

## 5. SYNERGIES AND COMPLEMENTARITIES

HRMOS occupies a unique position at R = 80,000 with 50 – 60 simultaneous targets, a parameter space entirely unoccupied by any existing or planned instrument on an 8-metre-class telescope. Its natural partners include Gaia[16], TESS, PLATO[17], and the Nancy Grace Roman Space Telescope[18], and the proposed Haydn mission (for target selection and candidate exoplanet follow up), and the forthcoming large MOS surveys with 4MOST, WEAVE, and MOONS (for synergistic low- and medium-resolution spectroscopy of the same stellar populations).

Looking ahead to the ELT era, HRMOS and ELT/ANDES are highly complementary: ANDES delivers ultimate sensitivity and resolution for individual targets, while HRMOS provides the statistical samples required to turn individual measurements into population studies. HRMOS also synergizes with advances in laboratory atomic physics and 3D non-LTE stellar atmosphere modelling, which are necessary to fully exploit the precision of R = 80,000 spectroscopy.

## 6. CONCLUSIONS

HRMOS represents a transformative capability for the ESO VLT: the first instrument to combine very high spectral resolution (R = 80,000) with genuine multi-object capability (50 – 60 simultaneous targets) and high-precision radial velocities (10 m s$^{-1}$). The design is well advanced, building on proven technology from MOONS, ESPRESSO, KMOS, and ANDES, and has been shown to meet all top-level requirements.

The science case spans a wide range of astrophysical priorities, Galactic exoplanet demographics, nucleosynthesis, Milky Way Satellite assembly, nucleocosmochronology, and star cluster physics, delivering breakthroughs not achievable with any existing or planned facility.

## ACKNOWLEDGEMENTS

The HRMOS project is the result of a collective effort of a large international consortium. The authors acknowledge all HRMOS consortium members; the full author list is given in the HRMOS White Paper v2.